\documentclass[12pt]{iopart}
\usepackage{graphicx} 

\newcommand{\EF}{E$_{\rm F}$ }
\newcommand{\SmFeAsOF}{SmFeAsO$_{1-x}$F$_{x}$ }
\begin{document}

\title[]{Evidence of the isoelectronic character of F doping in \SmFeAsOF: a first-principles investigation}

\author{Fabio Bernardini}

\address{Department of Physics, University of Cagliari, 09042 Monserrato, Italy.}

\author{Federico Caglieris}

\address{Leibniz-Institute for Solid State and Materials Research, 01069 Dresden, Germany.}

\author{Ilaria Pallecchi}

\address{CNR-SPIN, c/o Department of Physics, via Doedecaneso 33, 16146 Genova, Italy}

\author{Marina Putti}

\address{Department of Physics, University of Genova, Via Dodecaneso 33, 16146 Genova, Italy}

\ead{fabio.bernardini@dsf.unica.it}
\begin{abstract}
We study the electronic structure of the SmFeAsO$_{1-x}$F$_{x}$ alloy by means of first-principle calculations. We find that, contrary to common believe, F-doping does not change the charge balance between electrons and holes free-carriers in SmFeAsO$_{1-x}$F$_{x}$. For energies within a narrow energy range accross E$_{\rm F}$, the effect of F-doping on the band structure dispersion is tiny in both the paramagnetic and stripe antiferromagnetic phase. The charge balance between the conducting FeAs-layer and the SmO$_{1-x}$F$_{x}$ “charge reservoir” layer is not influenced by the compositional change. The additional charge carried by fluorine, with respect to the oxygen, is compensated by a change in the oxidation state of the Sm ion from 3+ to 2+. A comparison with the SmFe$_{1-x}$Co$_{x}$AsO system shows that such charge compensation by the Sm ion is not shared by donors substituting at the Fe site. 
\end{abstract}

\maketitle

\section{Introduction}
The much investigated dome-shaped phase diagram of non-conventional superconductors such as high-T$_{\rm C}$ cuprates and iron pnictides shows the ground state of the system as a function of carrier doping. Indeed, carrier doping is the primary control parameter that tunes the ground state of the parent compound from the antiferromagnetic (AFM) phase to the superconducting phase of the substituted compounds, culminating in the optimal doping value that maximizes the critical temperature T$_{\rm C}$. 

The {\it RE}FeAsO ({\it RE}=rare earth) family of iron pnictides, also called FeAs-1111 family, is the one with the largest transition temperature T$_{\rm C}$=58K at optimal doping \cite{1}. In these compounds, the superconducting transition is obtained by chemical substitution either in the conducting FeAs-layer that hosts Cooper pairs \cite{2} or in the so-called \textquotedblleft charge reservoir" layer. In the former case, Co substitution at the Fe site is used \cite{3}, while in the latter, superconductivity can be induced by oxygen vacancies \cite{4}, hydrogen or fluorine substitution of oxygen \cite{5}. The chemical doping quenches the AFM magnetic order and a superconducting phase is formed when the magnetic order disappears, even if, a narrow coexistence region can be present in the phase diagram of some compounds. Common believe suggests that the quenching of magnetic order in the FeAs-1111 family is a byproduct of electron doping, indeed, the above mentioned defects are donors \cite{6}. Notably, F (and Co) position in the periodic table is on the right of the element they substitute. At room temperature, the paramagnetic phase of the parent compounds of the FeAs-1111 family share a quasi-two dimensional electronic structure characterized by the presence of almost-nested electron and hole Fermi surfaces (FSs). The nesting is considered to be at the origin of the magnetic order. Any charge doping, by shifting the Fermi level across the bands, changes the shape of the FSs, disrupts the FSs nesting and quenches the magnetic order. 

This picture is somewhat challenged by the fact that isovalent doping (e.g. P substituting As) suppresses the AFM order as well and yields superconductivity \cite{8}. Therefore, electron doping is not strictly necessary to promote superconductivity in the FeAs-1111 family. Moreover, recent experiments show that the classification of F as a donating impurity is questioned \cite{9}. Flourine concentration reported in experimental works \cite{10,11,12,13,14}, is actually a nominal value, given that the actual content of the light element F is not easily measured. However, it is well known that the actual F content is lower than the nominal value, even if, with high-pressure synthesis methods severe F loss in gas-releasing FeAs-1111 compounds is prevented, as compared to solid-state reaction methods carried out at ambient pressure \cite{15}. Fluorine dopant is hardly incorporated into the structure in thin films deposited by pulsed laser ablation, even with excess F present in the ablation targets, yielding low T$_{\rm C}$ values, while in thin films deposited by molecular beam epitaxy high T$_{\rm C}$ values are obtained, indicating more effective F incorporation \cite{16,17}. These facts introduce uncertainties in comparing experimental phase diagrams. Even more importantly, it is crucial to clarify to what extent the actual F substitution is effective in introducing charge carriers (electrons) in the Fe-3$d$ bands crossing the Fermi level in FeAs-1111 compounds. Indeed, it has been recently evidenced that the charge carrier density introduced via F substitution in SmFeAsO and measured in unambiguous and precise way by fitting Shubnikov de Haas (SdH) oscillations, is significantly smaller than the nominal F doping, namely by a factor as large as 10 \cite{9}.
Therein, phase inhomogeneity was ruled out as sole possible reason for the discrepancy and it was tentatively suggested that F does not behave as a shallow donor but may be rather a deep impurity behaving as an isovalent substitutional impurity. Other works on F-doped FeAs-1111 report charge carrier density values measured by Hall effect compared with what expected from nominal F doping, namely significantly smaller values around 10\%-30\% are measured in ref.\cite{8}, in agreement with ref.\cite{16}, while comparable values in the range 40\%-200\% are reported in ref.\cite{10}. This puzzle requires to be sorted out by exploring how F substitution affects the band structure and the density of states over a wide energy spectrum, both at the Fermi level and well below.

In ref.\cite{19,20}, first-principles calculations of the band structures of FeAs-1111 pnictides having oxygen fully replaced by F, namely on SrFeAsF and CaFeAsF, show that in the vicinity of the Fermi level these compounds have essentially the same band dispersion as their counterparts with F fully replaced by oxygen. These results suggest that F-doping may act in a way more similar to an isoelectronic substitution rather than a donating impurity. However for iron fluoride systems, the F-2$p$ states are separated in energy from As-4$p$ ones in contrast to SrFeAsO, where O-2$p$ states strongly overlaps with As-4$p$. Moreover, Fermi surfaces of fluorides are found to be strongly two-dimensional. Such two-dimensional character has been confirmed by quantum oscillation measurements in CaFeAsF crystals, whose Fermi surface turns out to be formed by a pair of Dirac electron cylinders and a normal hole cylinder, with almost compensating carrier densities of 2.2$\cdot$10$^{19}$ and 2.6$\cdot$10$^{19}$ cm$^{-3}$, respectively. 

To shed light into the effectiveness of doping by F substitution in the FeAs-1111 family, in this work we present first-principles calculations of the band structures of F-doped SmFeAsO and compare those results with similar calculations made for Co-doped SmFeAsO. The comparison of the electronic structures clarifies the issue of doping effectiveness related to F substitution and provides precious information for the theoretical models addressing the pairing mechanism triggered by doping in these unconventional superconductors.

\section{Method}
The electronic structure is computed within the density functional theory (DFT) using the full-potential linearized augmented plane-wave (FLAPW) package Wien2k \cite{22} . We use the Perdew, Burke and Ernzerhof variant of the generalized gradient approximation to the exchange-correlation functional \cite{23}. We choose muffin-tin radii R$_{\rm MT}$ to be 2.4 $a_0$ for Sm, 2.3 $a_0$ for Fe and Co, 2.1 $a_0$ for As and 1.8 $a_0$  for O and F. The plane-wave cutoff is chosen to be R$_{\rm MT}$$\times$K$_{\rm max}$=9. Transition metal (TM) semicore 3$s$ and 3$p$, rare-earth 4$d$ states are explicitly taken into account as valence states. Magnetism is treated within the collinear formalism. Sm-4$f$ states are partially filled, within DFT this would bring the $f$-bands at the Fermi level, contrary to experimental finding, therefore, to account for the intratomic correlation effects a Hubbard term is added for Sm-4$f$ orbitals, the so-called DFT+U approach \cite{24}. The value for the Coulomb repulsion term U is chosen to be 9.7 eV, following our previous calculations \cite{25}. The Brillouin zone integration is performed by the modified tetrahedron method after Bl\"ochl implemetation \cite{26}. A $12\times12\times5$ Monkhorst-Pack mesh is used for the selfconsistent calculation, a  $64\times64\times12$ mesh is used to map the electronic structure and to draw the FSs.
We are interested in the effect of doping in both the paramagnetic and the AFM phase. The AFM phase is simulated using a stripe ordered structure, that comes out to be the ground state of SmFeAsO in DFT, in agreement with experiments. We use two kinds of supercell for those calculations, a 4 formula-unit cell for the AFM phase and a 2 formula-unit cell for the paramagnetic one. The former supercell has the basal plane edges rotated by 45$^\circ$ with respect to the latter. We use experimental data for the structures. For SmFeAsO we use $a$=$b$=3.9391Å, $c$= 8.497 \AA,  z$_{\rm Sm}$=0.1368 and z$_{\rm As}$=0.6609. Structural parameters of SmFeAsF are unknow because this compound is not thermodynamically stable and only up to 20\% F substitution is possible in real \SmFeAsOF samples, therefore we use for F-doped compounds the structural data of SmFeAsO. For SmCoAsO the effect of doping is very relevant, therefore any structural effect would not change our conclusions and we decided to keep for SmCoAsO the same structure we use for SmFeAsO. To study the charge doping effect of atom substitutions we performed an analysis of the Bader charges \cite{28}. The charges are computed by the {\it aim} post-processig tool included in the Wien2k package. According to Bader’s theory the crystalline primitive cell is divided into atomic sub-volumes. Symmetry-equivalent atoms will have same sub-volumes and total charges. The sum of the atomic charges (included the nuclei charges) must be zero in the primitive cell. Any charge transfer from atoms will be represented by positive atomic charge for the donating element (cation) and negative charge for the accepting element (anion). The charge balance between atomic layer can be studied by summing up the atomic charges of the atoms within each layer. This approach was successfully used in studying the effect of pressure of the charge state of  CaFeAsF and SrFeAsF compounds \cite{29}.   

\section{Electronic structure of SmFeAsO$_{1-x}$F$_x$: the paramagnetic phase. }
\begin{figure}[tb]
\centering{\includegraphics[width=0.9\textwidth]{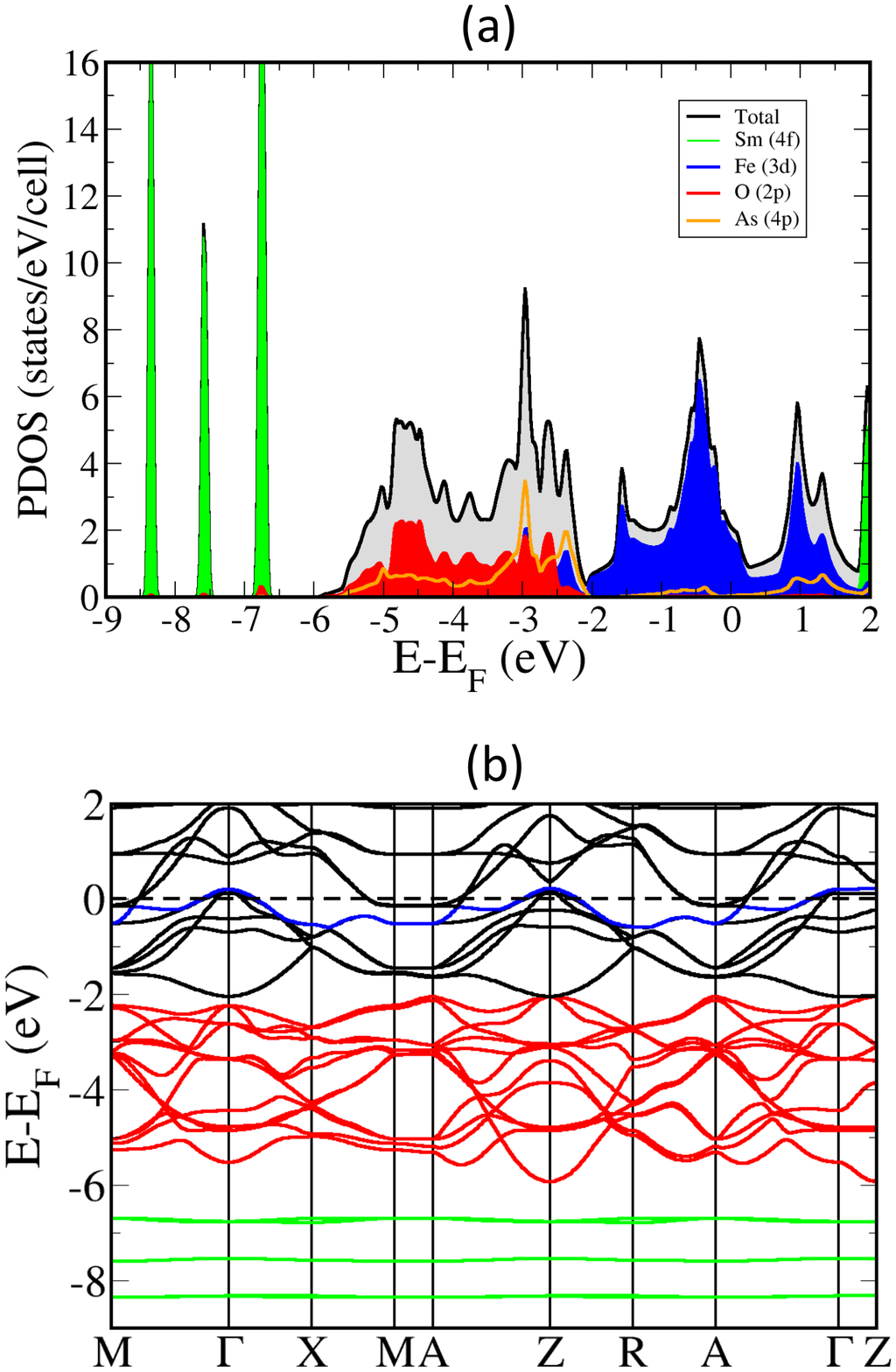}
}
\caption{PDOS (panel (a)) and band structure (panel (b)) for the SmFeAsO paramagnetic phase (DFT+U). Sm-4$f$ bands are colored in green, O-2$p$ in red, the uppermost hole-band in blue.}
\label{fig:pSmFeAsO}
\end{figure}

In Fig.\ref{fig:pSmFeAsO}(a) we show the projected density of states (PDOS) of SmFeAsO in the paramagnetic phase. We find that in a range from -2 to +2 eV (we set E$_{\rm F}$ as the reference energy) the density of states (DOS) is dominated by the contribution of Fe-3$d$ orbitals. From -6 to -2 eV we see a dominant contribution of O-2$p$ orbitals with a non-negligible component coming from the hybridization of Fe-3$d$ and As-4$p$ orbitals. Sm-4$f$ orbitals give rise to narrow bands located below -6 eV and above +2 eV. Fig.\ref{fig:pSmFeAsO}(b) shows the band structure of SmFeAsO along the high symmetry directions of the 2 formula-unit tetragonal cell. To help readability, we use a color legend for the band structure. We draw in green the bands related to Sm-4$f$ orbitals, in red those with a strong O-2$p$ character, in blue we highlight the uppermost band forming hole-FS. The bands structure resembles that of LaFeAsO \cite{30} and is in qualitative agreement with the ARPES measurement on Co-doped SmFeAsO \cite{31}.

\begin{figure}[tb]
\includegraphics[width=0.9\textwidth]{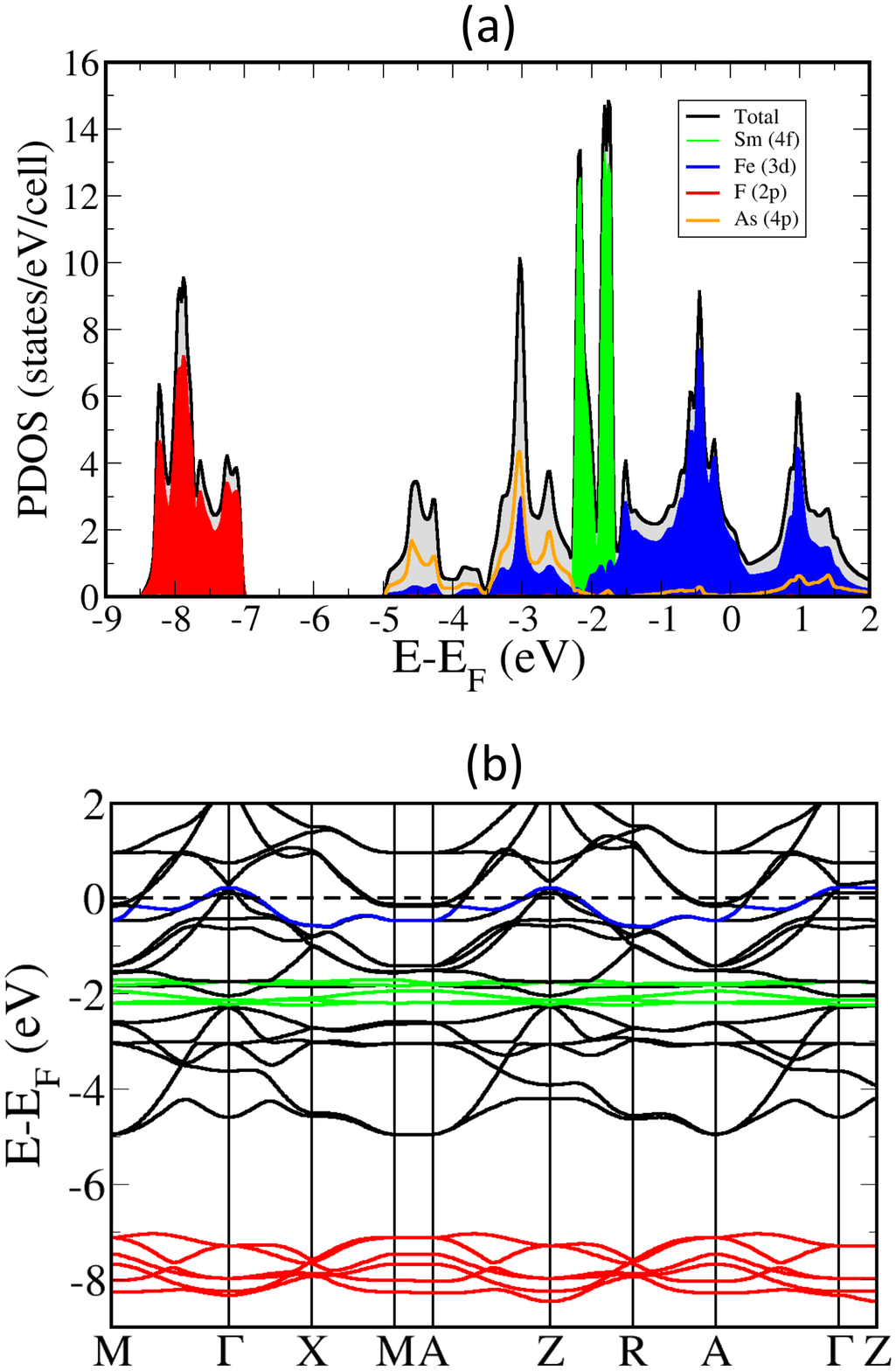}
\caption{PDOS (panel (a)) and band structure (panel (b)) for the SmFeAsF paramagnetic phase (DFT+U). Sm-4$f$  bands are colored in green, F-2$p$ in red, the uppermost hole-band in blue.}
\label{fig:pSmFeAsF}
\end{figure}

Simulation of F-doped SmFeAsO (SmFeAsO$_{1-x}$F$_x$) is technically more difficult. On one side low F-concentrations $x$ would require unfeasibly large supercells simulations, on the other a real random doping would disrupt the band structure that is a concept at the base of the present analysis. Experiments, such as the SdH oscillations, show that features related to the band structure concept persist even in doped SmFeAsO \cite{16}. Therefore, it is justified to study the SmFeAsO$_{1-x}$F$_x$ alloy as a system with the periodic structure of its parent compound.
We start our study of F-doping by comparing the band structure of SmFeAsF with that of SmFeAsO. In Fig.~\ref{fig:pSmFeAsF} we show the PDOS and the band structure of SmFeAsF in the 2 formula-unit cell. We draw in green the bands related to Sm-4$f$ orbitals, in red those with a strong F-2$p$ character, in blue the uppermost band forming hole-FS. We see a sizeable change in the total DOS. By a detailed inspection of the PDOS, we see that most of the difference between SmFeAsO and SmFeAsF is due to the position of F-2$p$ and Sm-4$f$ orbitals. The PDOS of Fe-3$d$ and As-4$p$ orbitals is only slightly affected by F substitution, mostly in the range from -6 to -2 eV. We note that the DOS around E$_{\rm F}$ does not seem to be modified by the compositional change. We see that the F-2$p$ bands are well below the bands related to the conducting FeAs layer. Notably, both compounds show the same dispersion of uppermost hole-band (highlighted in blue color).   
This finding can be rationalized looking at the oxidation states of O, F and Sm ions in those compounds. Oxygen oxidation state is -2, F is -1 and Sm can be +3 or +2. In the SmFeAsO compound the oxidation state of the Sm ion is +3. Sm looses three electrons, two in favour of the O$^{2-}$ ion and one to stabilize the FeAs layer. Instead, in the SmFeAsF compound, the Sm ion has a +2 oxidation state because it looses only two electrons, the first fills the F outer orbital, the other goes to the FeAs layer. This statement, crucial for our understanding of F-doping in Sm-1111, is supported by both the probability distribution inside the muffin-tin spheres, and the analysis of the Bader charges. The FLAPW method provides the orbital-decomposed probability distribution per muffin-tin sphere. For the 4$f$ orbital of the Sm atom  we obtain an average occupancy of 4.97 and 5.83 for SmFeAsO and SmFeAsF respectively.  Therefore, we conclude that in the SmFeAsO compound the Sm$^{3+}$ ion has a 4$f^5$6$s^0$ configuration, while in the SmFeAsF compound the Sm$^{2+}$ ion has a 4$f^6$6$s^0$ configuration. The difference in oxidation state is provided by a change in the occupation of the 4$f$ orbital. The different occupation of the Sm-4$f$ orbital does reflect on the position of the 4$f$ component of the PDOS. The larger is the occupation of an orbital, the higher is the effect of the inter-atomic Coulomb repulsion. In the SmFeAsO compound we find the Sm-4$f$ derived bands in a range between -8.5 and -6.5 eV while in the SmFeAsF the same states are found between -3 and -1.5 eV. The hybridization of the Sm-4$f$ orbitals with the Fe-3$d$ ones is very small, therefore the dispersion of the bands at the Fermi level is very similar in both compounds. The analysis of the Bader charges gives further support to our conclusion. The values of the charges for all the parent compounds studied in this work are given in Table \ref{tab:Bader}. The sum of the Sm (Q$_{\rm Sm}$) and O (or F) (Q$_{\rm O/F}$) charges (electrons+nuclei) provides the amount of charge that is transferred from the “reservoir” in to the “conducting” layer. We see that Q$_{\rm Sm}$+Q$_{\rm O/F}$ does not change with composition within an uncertainty of 0.01 elementary charges. F is a much less negative ion than O (Q$_{\rm F}-{\rm Q}_{\rm O}$=0.497) but this difference is {\it exactly} canceled by the change in the charge status of Sm. Therefore, F-doping does not change the overall charge balance between \textquotedblleft reservoir" and \textquotedblleft conducting" layers. As we will see below, doping with Co that substitutes Fe does not change the charge balance between the above mentioned layers either, meaning that the extra electron charge given by the cobalt atom is kept inside the “conducting” layer. 

{\begin{table}[h] 
\small
\caption{Values of the atomic charges (electrons+nucleus) according to Bader’s theory. Values are computed in the paramagnetic phase.} 
\centering{
\begin{tabular}{lccc}
\hline \hline
     & SmFeAsO	& SmFeAsF	& SmCoAsO  \\
\hline 
Q$_{\rm Sm}$   & +1.945	    &+1.457	&+1.945 \\
Q$_{\rm Fe/Co}$& +0.144	    &+0.171	&+0.045 \\
Q$_{\rm As}$   & -0.768	    &-0.808 &-0.668 \\
Q$_{\rm O/F}$  & -1.299	    &-0.802	&-1.299 \\
Q$_{\rm Sm}+
$Q$_{\rm O/F}$ & +0.646	    &+0.655 &+0.646 \\
\hline \hline 
\end{tabular}}

\label{tab:Bader}
\end{table}}

\begin{figure}[tb]
\includegraphics[width=\textwidth]{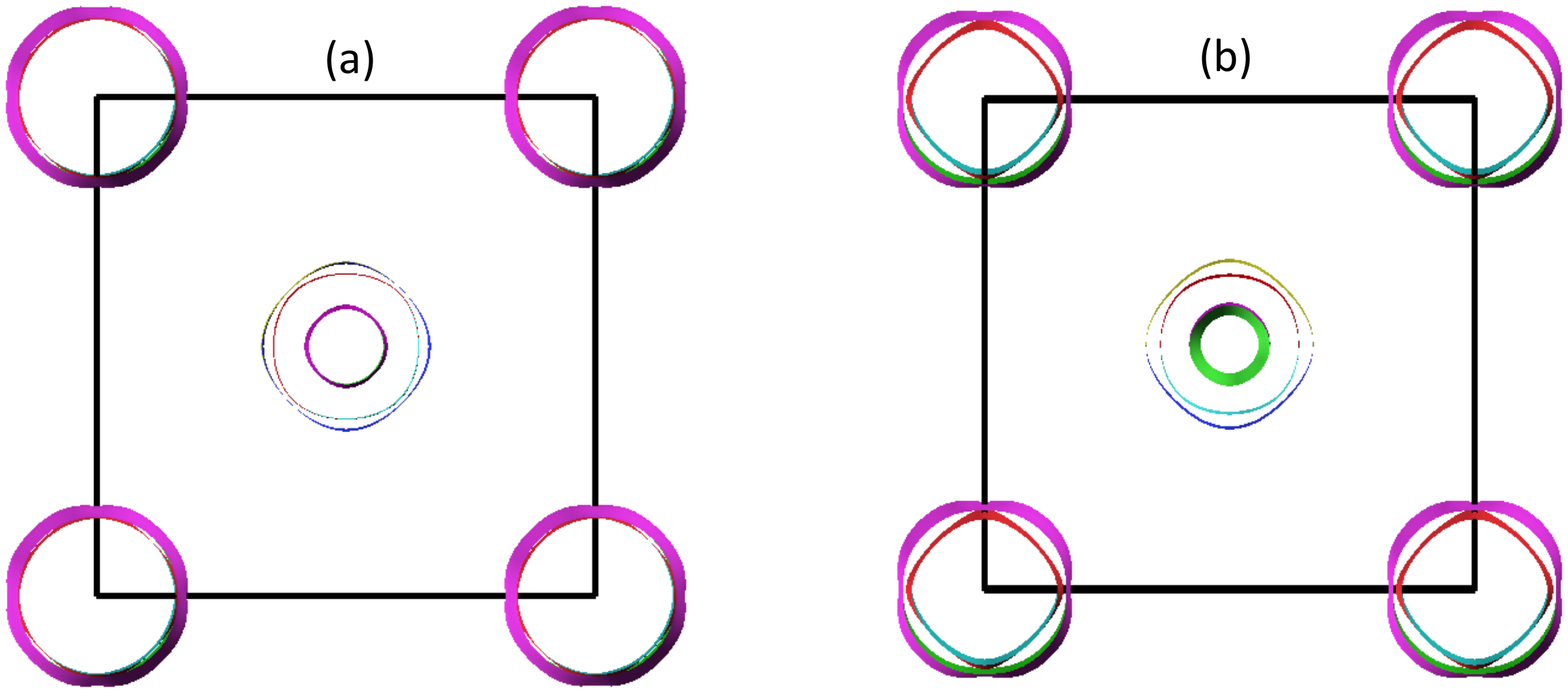}
\caption{Fermi surface of SmFeAsO (a) and SmFeAsF (b) in the paramagnetic phase.}
\label{fig:FS-SmFeAsX}
\end{figure}

In Fig.\ref{fig:FS-SmFeAsX}, we compare the FSs of SmFeAsO with those of SmFeAsF. As usual, in these compounds we have five FS sheets. Three hole-FSs at the Brillouin zone (BZ) center and two electron-FSs at the corners of the BZ.  The inner hole-FS is the only FS sheet showing a sizable difference between the compounds. It has a larger dispersion along the c-axis direction in the SmFeAsF with respect to SmFeAsO. The second and third hole-FSs show the same size and shape in both compounds. As for the electron-FSs, the inner one has a bit more rounded shape in SmFeAsO than in SmFeAsF. 

\begin{figure}[tb!]
\includegraphics[width=\textwidth]{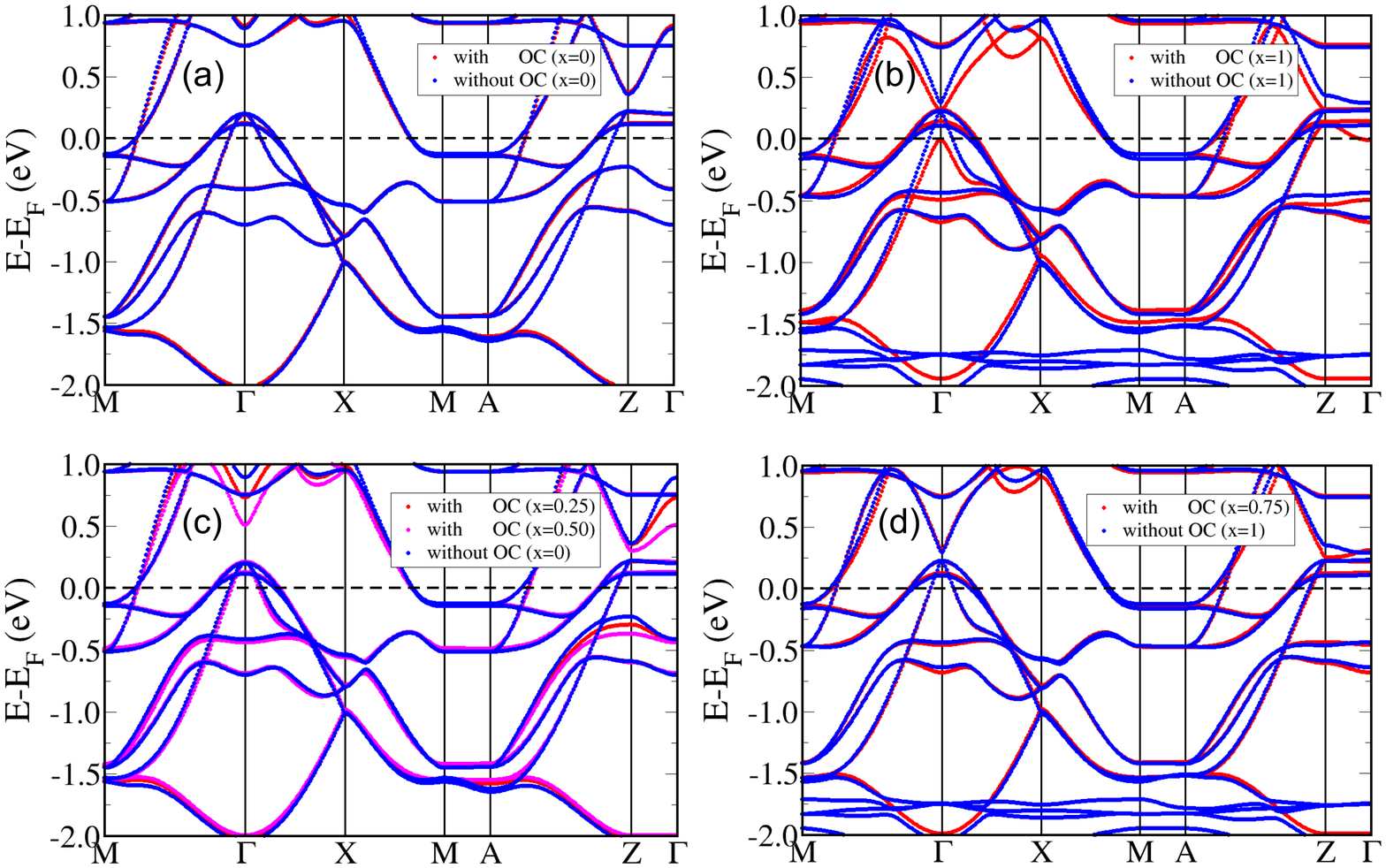}
\caption{Band structure of the SmFeAsO$_{1-x}$F$_x$ alloy in the paramagnetic phase. Panel (a) and (b):  results obtained with (red line) and without (blue line) the OC approximation for $x$=0 and 1.  Panel (c): bands for $x$=0.25 (red) and 0.50 (pink) with OC, compared with the bands of SmFeAsO without OC (blue). Panel (d): bands for $x$=0.75 (red) with OC, compared with the bands of SmFeAsF without OC (blue).}
\label{fig:vcaoc}
\end{figure}

So far we have come to the conclusion that, in the paramagnetic phase of Sm-1111, the full substitution of O with F does not lead to any relevant change in the band structure near the Fermi level. This finding does not guarantee that the band structure of the SmFeAsO$_{1-x}$F$_x$ alloy has the same band dispersion and relative Fermi level position as its two stoichiometric endpoints. To shed light on this issue, we compute the band structure for the SmFeAsO$_{1-x}$F$_x$ alloy at different concentrations. A simple and  common way to study alloys is the so-called Virtual Crystal Approximation  (VCA) \cite{32}. With this approach the SmFeAsO$_{1-x}$F$_x$ alloy is simulated in the usual 2 formula-unit cell by a virtual atom of nuclear charge Z=8+$x$ placed at the O-site. The use of the VCA to study Sm-1111 poses an additional technical problem due to the use of the DFT+U approach in our calculations. Localized orbitals, such as Sm-4$f$, have the tendency to an integer occupation. This is in accordance with the DFT+U method, where a penalty functional is used to force integer occupation of a given localized orbital to mimic the effect of intra-atomic Coulomb interactions. While successful in describing stoichiometric compounds, this approach has a limit in coping with the alloys. Indeed, the SmFeAsO$_{1-x}$F$_x$ alloy should be considered a mix of Sm$^{3+}$ and      Sm$^{2+}$ ions with an effective chemical formula (Sm$^{3+}$)$_{1-x}$(Sm$^{2+}$)$_x$FeAsO$_{1-x}$F$_x$. Within the VCA, the Sm atom should be represented by a virtual ion with a 4$f^{5+x}$6$s^0$ configuration, that is, a configuration having a non integer occupation of the 4$f$ orbitals. This is not compatible with the DFT+U approach that forces integer occupations. To overcome this problem, we resort to the “open-core”  (OC) technique \cite{33}, a method used in all-electron calculation of rare-earth compounds to constrain a given oxidation state of a rare-earth ion. The 4$f$ orbitals are removed from the valence states and treated as if they were core states with fixed orbital occupation. Within the VCA+OC  the (Sm$^{3+}$)$_{1-x}$(Sm$^{2+}$)$_x$ virtual ion of the SmFeAsO$_{1-x}$F$_x$ alloy will have 5+$x$ occupation on the Sm-4$f$ orbitals. The results of the VCA+OC calculation of the electronic structure are shown in Fig.\ref{fig:vcaoc}, in a narrow range of energy around E$_{\rm F}$. First, we validate the OC approach comparing the band structures of SmFeAsO and SmFeAsF computed with and without the OC approximation (Figs. \ref{fig:vcaoc}(a) and (b)). For the SmFeAsO bands the agreement is perfect. Since we are interested in low F concentrations, this is a very important result. As for SmFeAsF, we see that in the OC  calculation the inner hole-band is a bit lower in energy at the zone center. In spite of this, the dispersion of the other bands forming the FSs and the relative position of E$_{\rm F}$ is very well reproduced. In Figs. \ref{fig:vcaoc}(c) and (d) we compare the band structure of the SmFeAsO$_{1-x}$F$_x$ alloy obtained by the VCA+OC with the band structure of the parent compounds. The only relevant effect of doping in the band dispersion is seen in the Z-$\Gamma$ direction. There, an empty electron-band lowers with the increase of F concentration. 


We know that the VCA has some limitations that we will discuss now. In the VCA, we use the same virtual atom to simulate both F and O. Therefore, the effect of charge transfer between F and O on the band structure is absent in VCA calculation. Moreover, the alloy can have an ordered super-structure. Alloy ordering can sometimes lead to sizeable effects on the alloy band dispersion. To double check the result obtained with the VCA+OC approach and show that an ordered alloy has a band structure very similar to the band obtained by the VCA+OC, we make an additional calculation of the electronic structure combining the supercell approach and the OC technique.  
We use a 4 formula-unit cell with three O atoms and one F atom. The Sm atoms are represented by three “open-core” Sm$^{3+}$ ions with a 4$f^5$6$s^0$ configuration and one “open-core” Sm$^{2+}$ ion with a 4$f^6$6$s^0$ configuration. We find that the ground state of the alloy does not show any magnetic moment on Fe. This result confirms that the DFT is able to predict the quenching of the magnetic order due to F-doping as found in the experiments.  

\begin{figure}[tb!]
\includegraphics[width=\textwidth]{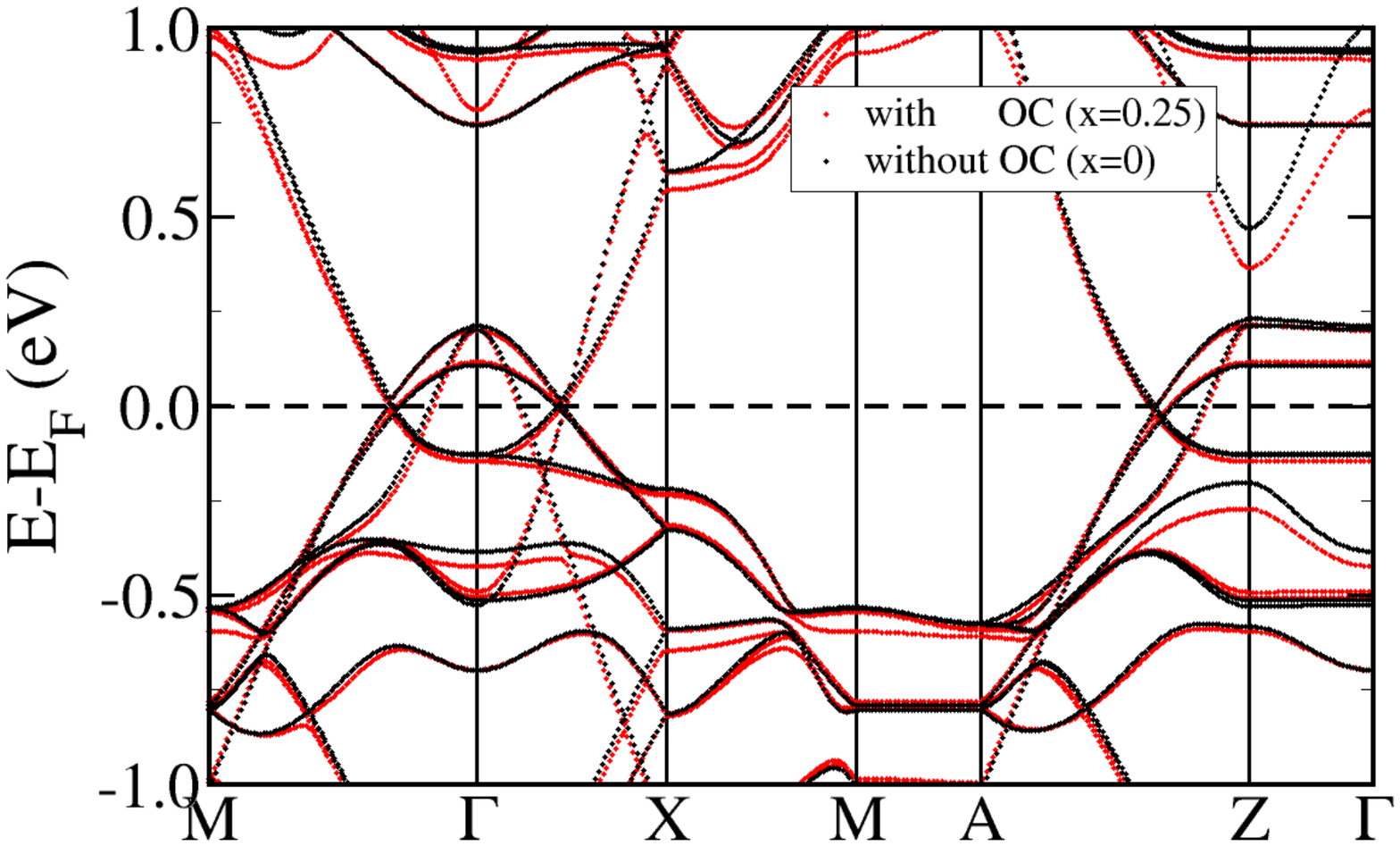}
\caption{Band structure of the SmFeAsO$_{1-x}$F$_x$ ordered alloy in the paramagnetic phase at $x$=0.25 (red dots). In the band structure plot, a comparison is made with the band structure of SmFeAsO without OC (black dots).}
\label{fig:ordered}
\end{figure}

In Fig. \ref{fig:ordered}, we compare the band structure of the SmFeAsO$_{0.75}$F$_{0.25}$ ordered alloy with that of the stoichiometric compound SmFeAsO. Since the band plot depends on the choice of the periodic cell we compared the alloy bands with those of the stoichiometric compound performed for the same supercell. In Fig. \ref{fig:ordered} we plot the bands in a narrow range of energy around the Fermi energy to observe the effect of doping on the band dispersion for low energies range of excitation. We see that the band dispersion around E$_{\rm F}$ does not change very much upon doping even in the case of an ordered alloy, confirming our conclusion that VCA+OC is a reliable approach to study the band structure of the \SmFeAsOF alloy.  
Summarizing, we can say that our calculations suggest that in the SmFeAsO$_{1-x}$F$_x$ alloy the band structure at E$_{\rm F}$ and the concentration of free carriers (holes and electrons) is not influenced by the concentration of F. Therefore, we conclude that, in the paramagnetic phase, the F atoms that substitute O in Sm-1111 behaves like a very deep donor (or an isovalent) substitutional defect does. The additional electron carried by the F atom is kept inside the SmX (X=O,F) \textquotedblleft charge reservoir" layer.    

\section{Electronic structure of Co-doped SmFeAsO: the paramagnetic phase}

\begin{figure}[tb!]
\includegraphics[width=0.9\textwidth]{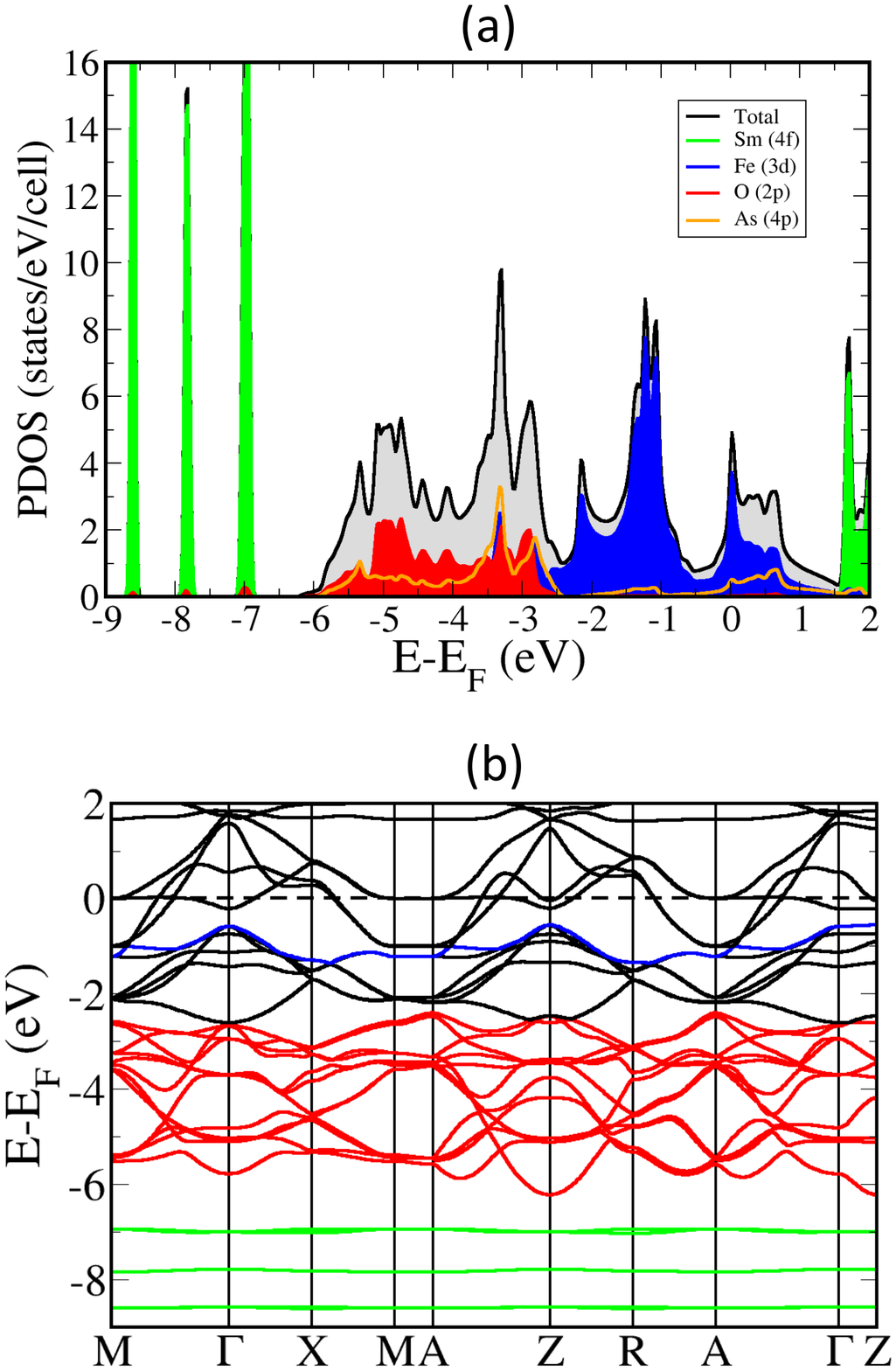}
\caption{PDOS (panel (a)) and band structure (panel (b)) of SmCoAsO in the paramagnetic phase.
Sm-4$f$  bands are colored in green, O-2$p$ in red, in blue the band that in SmFeAsO was the uppermost hole-band.}
\label{fig:pSmCoAsO}
\end{figure}

We want now to show that the deep donor (or isovalent) behavior that we find for F substituting at O is not a general feature shared by  every dopants in SmFeAsO. For instance, we find a completely different behavior in the SmFe$_{1-x}$Co$_x$AsO alloy, where Co substitutes Fe. In Fig. \ref{fig:pSmCoAsO} we show the PDOS and the band structure for the paramagnetic phase of SmCoAsO. The shape of the PDOS of SmCoAsO is very similar to that of SmFeAsO. The most important difference between the compounds is the position of E$_{\rm F}$.

In the SmFeAsO, E$_{\rm F}$ falls in the pseudogap of the transition metal 3$d$ bands, while, in the SmCoAsO, E$_{\rm F}$ coincides with a peak in the TM-3$d$ component of the PDOS. Besides the change in the position of E$_{\rm F}$, the band dispersion shown in Fig. \ref{fig:pSmCoAsO}(b) resembles that of Fig. \ref{fig:pSmFeAsO}(b). We notice that the energy of the bands originated by the Sm-4$f$ orbitals is unchanged upon Co-doping. This means that the oxidation state of the Sm ion (Sm$^{3+}$) is the same in SmFeAsO and SmCoAsO. Indeed, the calculated occupancy of the Sm-4$f$ orbitals in SmCoAsO is 4.97, the same found in SmFeAsO and the Bader charge for Sm (Q$_{\rm Sm}$=+1.945) is the same in both compounds, within numerical accuracy. The Co atom has an additional valence electron as compared to the Fe atom. It follows that in SmCoAsO, the band manifold related to the TM-3$d$ orbitals contains one electron more than in SmFeAsO, and, as a consequence, E$_{\rm F}$ is raised. Moving up the Fermi level has a great effect on the FSs shape that looses the remarkable nesting among the electron- and hole-FSs. The nesting is at the origin of the AFM order, loosing the nesting meas that other kind of magnetic orders (e.g. the FM) may be favored. We know that superconductivity in the FeAs-1111 require the system to be a stripe AFM before the chemical doping trigger the superconductivity. Therefore, it is not surprising that Co is less efficient than F as dopant for the superconducting behavior. Experiments found for LaFe$_{1-x}$Co$_x$AsO a quite narrow range for the doping concentration sustaining a superconducting behavior \cite{3}. At high Co concentration the LaFe$_{1-x}$Co$_x$AsO system ceases to be a superconductor and turns out to be ferromagnetic material.
%

\section{Electronic structure of SmFeAsO$_{1-x}$F$_x$: the stripe antiferromagnetic  phase}

Our conclusions about the effect of F-doping on the electronic structure of Sm-1111 are based on DFT calculations. Those calculations are confirmed by experimental evidence in the case of the stripe-AFM (s-AFM) phase for the F-doped SmFeAsO. Indeed, originally the present work was stimulated by a recent experimental work on the Shubnikov-de Haas (SdH) oscillation of the \SmFeAsOF alloy \cite{16}. In the latter work, the authors pointed out the very small influence that F-doping has on the SdH oscillations at low temperature. This work is especially significant because, in the s-AFM phase, the cross-sectional areas of the FSs are very small and any effect of charge doping on the FSs cross-sectional area would be detected easily. With respect to the case of the paramagnetic phase, a theoretical prediction of the effect of F-doping on the electronic structure for the s-AFM phase is more difficult from the technical point of view, because we cannot simulate a very low doping concentration with a supercell approach and an $x$=0.25 doping concentration makes the system already paramagnetic. Therefore, we will limit our study of the s-AFM phase to the stoichiometric endpoints of the SmFeAsO$_{1-x}$F$_x$ alloy. Our aim is to show that, as it happens in the paramagnetic phase, for the s-AFM too the electronic structure for $x$=0 and $x$=1 are similar, and there is no reason to believe that, for intermediate values of $x$, the electronic structure would be strongly dependent on doping. The magnetic phase is present a small change of the lattice parameters breaking the tetragonal symmetry. In this work we overlook this structural effect. The SmFeAsF compound is not thermodynamically stable and its structure is unknown. Therefore, small effect such as tetragonal vs. orthorhombic structure distortion are irrelevant in the context of a comparative study of SmFeAsO and SmFeAsF in the magnetic phase. An important feature of the electronic structure of stoichiometric SmFeAsF is that it has a stable s-AFM phase. We strongly believe that this is not an artifact of the DFT calculation. Indeed, there is experimental evidence that a similar compound, the LaFeAsO$_{1-x}$H$_x$, has a second AFM phase at high $x$ values \cite{37}. 

\begin{figure}[tb!]
\includegraphics[width=0.9\textwidth]{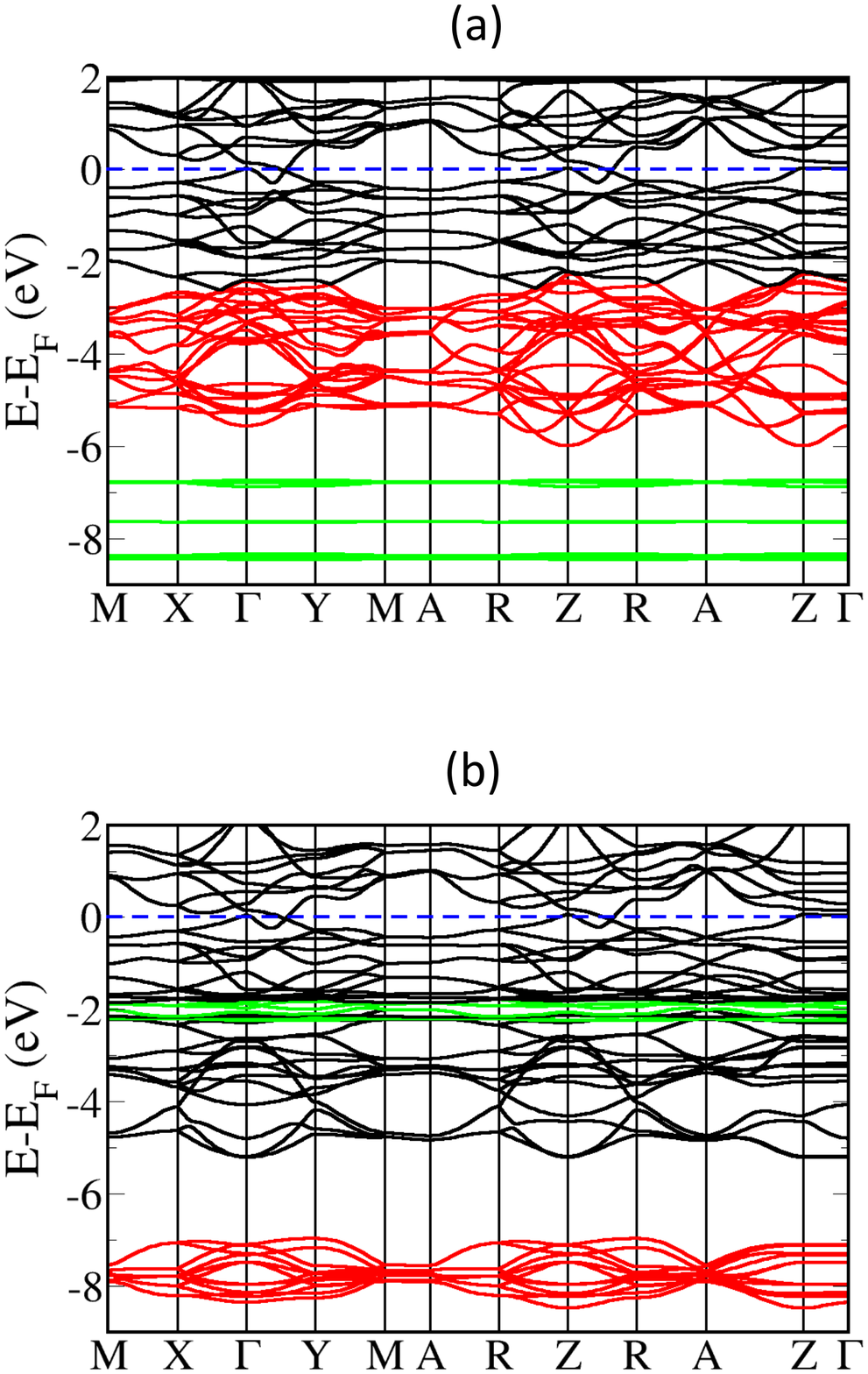}
\caption{Band structure of SmFeAsO (panel (a)) and SmCoAsO (panel (b)) in the s-AFM phase. Sm-4$f$  bands are colored in green, O-2$p$ and F-2$p$ in red.}
\label{fig:AFMbands}
\end{figure}

In Fig. \ref{fig:AFMbands} we compare the band structure of the two stoichiometric endpoints of the SmFeAsO$_{1-x}$F$_x$ alloy in the s-AFM phase. 
The band structures shown in Fig. \ref{fig:AFMbands} resemble that of Fig. \ref{fig:pSmFeAsO}(b) and \ref{fig:pSmFeAsF}(b), if we consider that the high-symmetry directions of the s-AFM phase are different with respect to those of the paramagnetic phase because the paramagnetic phase has a tetragonal symmetry while the s-AFM phase has an orthorhombic symmetry. 
We point out that, the change in the position of the band  related to the Sm-4$f$ orbitals (drawn in green), is the same as found for the paramagnetic phase. Therefore, in the s-AFM phase, we have the same change in the oxidation state of the Sm ion that was found in the paramagnetic phase.  No additional charge doping of the FeAs layer is provided and the position of \EF does not change. The analysis of the Bader’s charges further confirms that the total amount of charge on the FeAs-layer is the same in both compounds in the s-AFM phase.

\begin{figure}[tb!]
\includegraphics[width=0.9\textwidth]{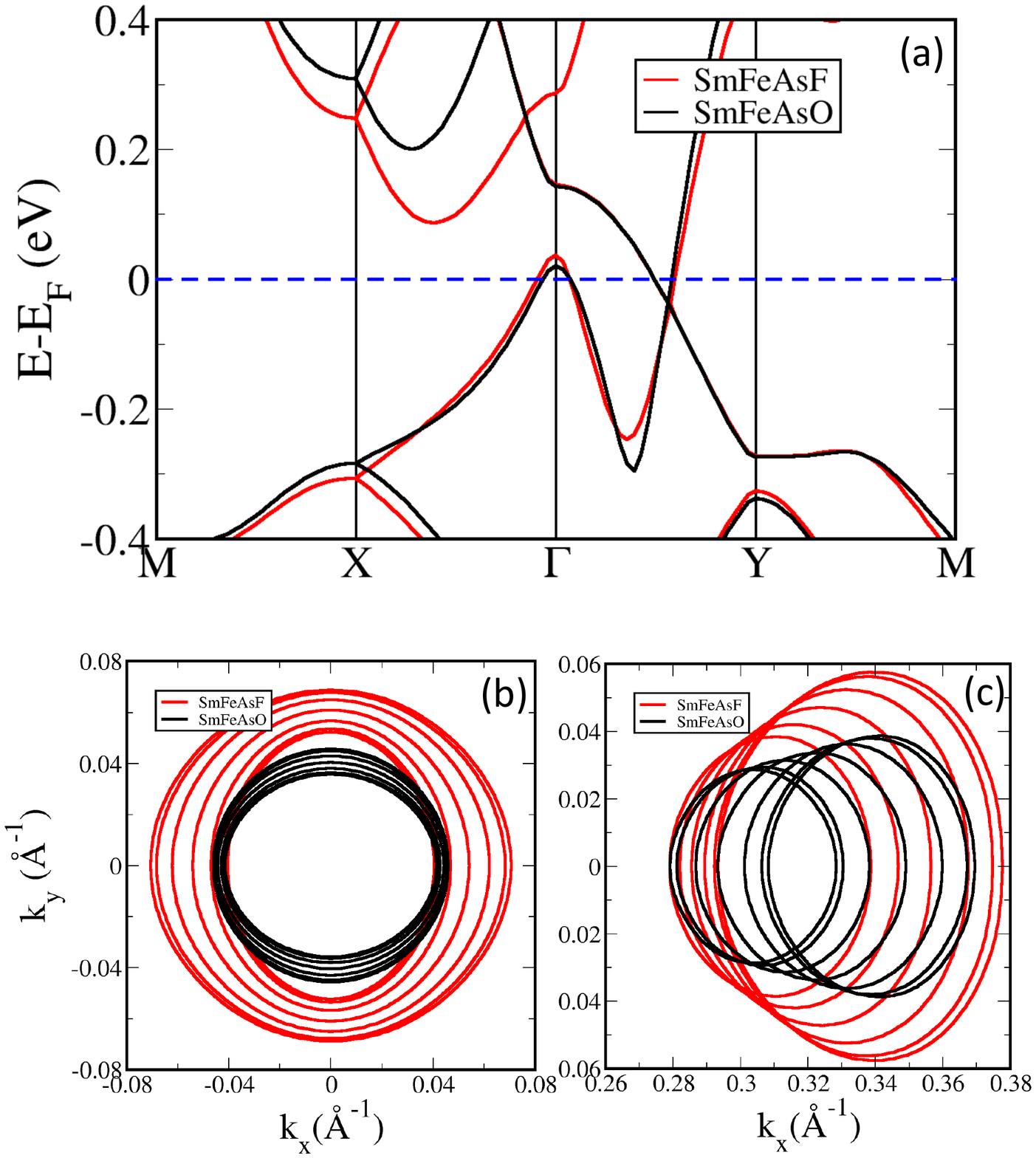}
\caption{Electronic structure of SmFeAsO and SmFeAsF in the s-AFM phase. Panel (a): Band structure dispersion around \EF. Panel (b) and (c): FSs cross sections at $k_{\rm z}$=const, for holes and electrons respectively.}
\label{fig:AFM-FS}
\end{figure}

The most relevant difference between the bands structures of the s-AFM and paramagnetic phase in both SmFeAsO and SmFeAsF is found in a narrow range across the Fermi level. There, a nearly full gap opens along the (0,$\pi$) direction, a contact points between empty and filled bands remains and leads to the formation of two Dirac cones (DCs). In Fig. \ref{fig:AFM-FS}(a) we compare the band dispersion of SmFeAsO and SmFeAsF around the DCs, while in Fig. \ref{fig:AFM-FS}(b) we compare the cross sections of the FSs. We have an hole-cylinder at the Brillouin zone center and two equivalent electron cylinders nearly midway $\Gamma$ and (0,$\pi$). 

In the s-AFM phase, the shape of the hole- and electron-FSs are slightly warped cylinders and both compounds have very small band dispersion along the $c$-axis direction. SmFeAsO is a compensated semimetal, therefore the sum of the volumes of the two electron-FSs must be equal to that of the single hole-FS. Electron- and hole-bands comes primarily from Fe-3$d$ orbitals. If the change in the valence charge, due to the substitution of O by F, is compensated by a different oxidation state of the Sm ion, the charge balance between electron and holes would be not modified. This picture is confirmed by the comparison of the FS cross-sections in Fig. \ref{fig:AFM-FS}(b). We note that going from SmFeAsO to SmFeAsF the cross-sectional area of both hole- and electron-band FSs increases, but the compensation between holes and electrons is kept. This finding confirms that F-doping does not induce an overall change of charge in the FeAs-layer. The effect of F-doping is a slight increase of the overlap between the electron- and hole-bands due to a small up shift of the hole-bands and a concurrent down shift of the electron-bands. It is worth to mention the similarity of the band structure we are discussing with the experimental results obtained recently for CaFeAsF \cite{21}. CaFeAsF and SmFeAsF share the same structure, the same magnetic order (s-AFM) and the same oxidation state for atoms at equivalent sites. Ca and Sm have a +2 oxidation state, we expect that the actual band structures of those compounds to be very similar.  In our recent work on the SdH oscillation in F-doped SmFeAsO  \cite{16}, we where not able to disentangle the contributions of hole- and electron-FSs, because of the absence of data on the angular dependence of the frequencies.  This does not mean that the data on the SdH oscillation are not compatible with the picture we are drawing here, that is the presence of a single hole-FS and two electron-FSs leading to full compensiation of carriers, with the electron-FSs coming out of the intersection of DCs with the Fermi level. 

\section{Summary and Conclusions}
In this work we study the compositional dependence of the electronic structure of the \SmFeAsOF alloy by first-principle methods using the VCA+OC approach.  
Our study suggests that the onset of superconductivity in this system is not due to a charge doping effect where additional charge is poured in to the conducting FeAs-layer by F atoms acting as donating defects. The additional charge provided by the F defect is kept inside the “reservoir” layer and does not participate to conduction because, in the \SmFeAsOF alloy, the donating effect of F substitution is compensated by the change in the oxidation state from +3 to +2 of a neighboring Sm ion. Since the bands with a strong Sm-4$f$ and F-2$p$ character are far below or far above the Fermi level, chemical doping does not disrupt the charge compensation between electrons and holes present in the SmFeAsO parent compound. The effect of F-doping on the band structure dispersion around \EF is really tiny, because those states are formed by the Fe-3$d$ and As-4$p$ orbitals whose average occupation is not affected by F, so that, from the point of view of free carriers, F behaves like a very deep donor or isovalent impurity.

F-doping induces superconductivity because it quenches the s-AFM order.  The transition between the s-AFM to the paramagnetic phase changes dramatically the topology of the FSs leading to the Lifshitz transition referred to in literature \cite{39}. In the s-AFM phase the need for compensation between one hole-FS and two electron-FSs does not allow the nesting of the FSs. Electron- and hole-FSs will always have cross-sections with 1:2 ratio. The nesting between electron- and hole-FSs is fundamental for the onset of the superconducting phase with $s\pm$ symmetry. Therefore, superconductivity with that symmetry is not compatible with the nearly full gapped band structure of the s-AFM phase. Our calculations show that the paramagnetic (and superconducting phase) has a nested FS structure. We want to stress that a nested structure is not incompatible with a paramagnetic phase as pointed out by Johannes and Mazin \cite{40}. In the iron pnictides, magnetism does not follow neither the itinerant nor the localized behavior, therefore a chemical doping, that does not modify the topology of the FS can still change the magnetic behavior of the material. The coexistence of nesting and paramagnetic phase, with large spin fluctuation but absence of static magnetism, is a relevant point to explain why F is so effective for the superconducting behavior despite its negligible effect on the charge state of Fe and As.  The same mechanism may also explain why isoelectronic P substitution of As makes LaFeAsO a superconductor \cite{41}.
Doping with Co has a very different effect on the Sm-1111 with respect to F. Substitution of Fe with Co does not affect the oxidation state of Sm. Doping by Co atoms is found to enrich the FeAs-layer charge. Each Co atom donates an electron to the TM-3$d$ bands. The donation raises E$_{\rm F}$, increases the DOS at E$_{\rm F}$, it destroys the FS nesting and favors, at high concentrations, the FM order over the s-AFM one.  
This finding can explain why F is a more efficient trigger for the onset of superconductivity with respect to Co and why high concentrations of F can sustain the superconducting phase while superconductivity is suppressed by high Co concentrations.

\section{Acknowledgments}
We acknowledge the support of FP7 European project SUPER-IRON (Grant Agreement No. 283204) and the support of the HFML-RU/FOM, member of the European Magnetic Field Laboratory (EMFL). FB acknowledges partial support from the project “Multiphysics approach to thermoelectricity” funded by Fondazione Sardegna (2017) and project PRID 2016 funded by the University of Cagliari. 


\end{document}